\documentclass[%
 reprint,
 amsmath,
 amssymb,
 aps,
 superscriptaddress,
 floatfix,
]{revtex4-1}

\usepackage{color}
\usepackage{siunitx}
\usepackage{graphicx}% Include figure files
\usepackage{braket}
\usepackage{verbatim}
\usepackage{dcolumn}% Align table columns on decimal point
\usepackage{bm}% bold math
\usepackage{hyperref}% add hypertext capabilities
\usepackage[main=english,vietnamese]{babel}

\newcommand{\mytitle}{Umklapp electron-electron scattering in bilayer graphene moir\'e superlattice}

\newcommand{\myaffiliations}{
\author{Christian Moulsdale}
\email{christian.moulsdale@postgrad.manchester.ac.uk}
\affiliation{\noindent School of Physics and Astronomy, University of Manchester, Manchester M13 9PL, UK}%
\affiliation{\noindent National Graphene Institute, University of Manchester, Manchester M13 9PL, UK}% 

\author{\noindent Vladimir Fal'ko}
\affiliation{\noindent School of Physics and Astronomy, University of Manchester, Manchester M13 9PL, UK}%
\affiliation{\noindent National Graphene Institute, University of Manchester, Manchester M13 9PL, UK}%
\affiliation{\noindent Henry Royce Institute, Institute for Advanced Materials, Manchester M13 9PL, UK}%}
}

\newcommand{\bilayerHamiltonian}{PhysRevLett.96.086805,McCann_2013}
\newcommand{\bilayerHamiltonianParameters}{PhysRevB.80.165406}
\newcommand{\ehScattering}{doi:10.1126/sciadv.abi8481}
\newcommand{\electronElectronLinearTransport}{Boltzmann,Boltzmann2,Wallbank2019}

\newcommand{\experimentMsl}{Yankowitz2012,Dean2013,Ponomarenko2013,doi:10.1126/science.aal3357,Wallbank2019}
\newcommand{\hbnDielectricConstant}{doi:10.1021/acs.nanolett.1c02211}

\newcommand{\mSL}{PhysRevB.87.245408,doi:10.1126/science.1237240,PhysRevB.94.045442,Yankowitz2019}
\newcommand{\minivalley}{PhysRevB.98.155435}
\newcommand{\mslPotentials}{doi:10.1126/science.aaf1095}

\newcommand{\screening}{PhysRevB.80.241402,PhysRevB.84.085112,doi:10.1021/acs.nanolett.1c02211,doi:10.1126/science.abb8754}

\newcommand{\trigonal}{PhysRevLett.127.046801}
\newcommand{\tunableBandgap}{PhysRevLett.102.256405,Zhang2009,doi:10.1021/nl1039499,doi:10.1021/nn202463g}
\newcommand{\twistedBilayerGraphene}{PhysRevLett.99.256802,Li2010,Cao2018,Cao2018_2}
\newcommand{\ueeMonolayer}{Wallbank2019,Kim2020}
\newcommand{\ueeQuadraticTemperature}{PhysRevB.23.2718}

\newcommand{\vanHoveResistivity}{PhysRevB.53.11344,Xu2021}
\newcommand{\zoneFolding}{PhysRevB.87.245408,PhysRevB.94.045442}

\begin{document}

\preprint{APS/123-QED}

\title{\mytitle} 

\myaffiliations

\date{\today}% It is always \today, today,
             %  but any date may be explicitly specified

\begin{abstract}
Recent experimental advances have been marked by the observations of ballistic electron transport in moir\'e superlattices in highly aligned heterostructures of graphene and hexagonal boron nitride (hBN). Here, we predict that a high-quality graphene bilayer aligned with an hBN substrate features $T^2$-dependent resistivity caused by umklapp electron-electron (Uee) scattering from the moir\'e superlattice, that is, a momentum kick by Bragg scattering experienced by a pair of electrons. Substantial Uee scattering appears upon $p$-doping of the bilayer above a threshold density, which depends on the twist angle between graphene and hBN, and its contribution towards the resistivity grows rapidly with hole density until it reaches a peak value, whose amplitude changes non-monotonically with the superlattice period. We also analyse the influence of an electrostatically induced bandgap in the bilayer and trigonal warping it enhances in the electron dispersion on the electron-electron umklapp scattering.
\end{abstract}

\maketitle

\begin{figure}[ht!]
\centering
\vspace{10pt}
\includegraphics[width=\linewidth]{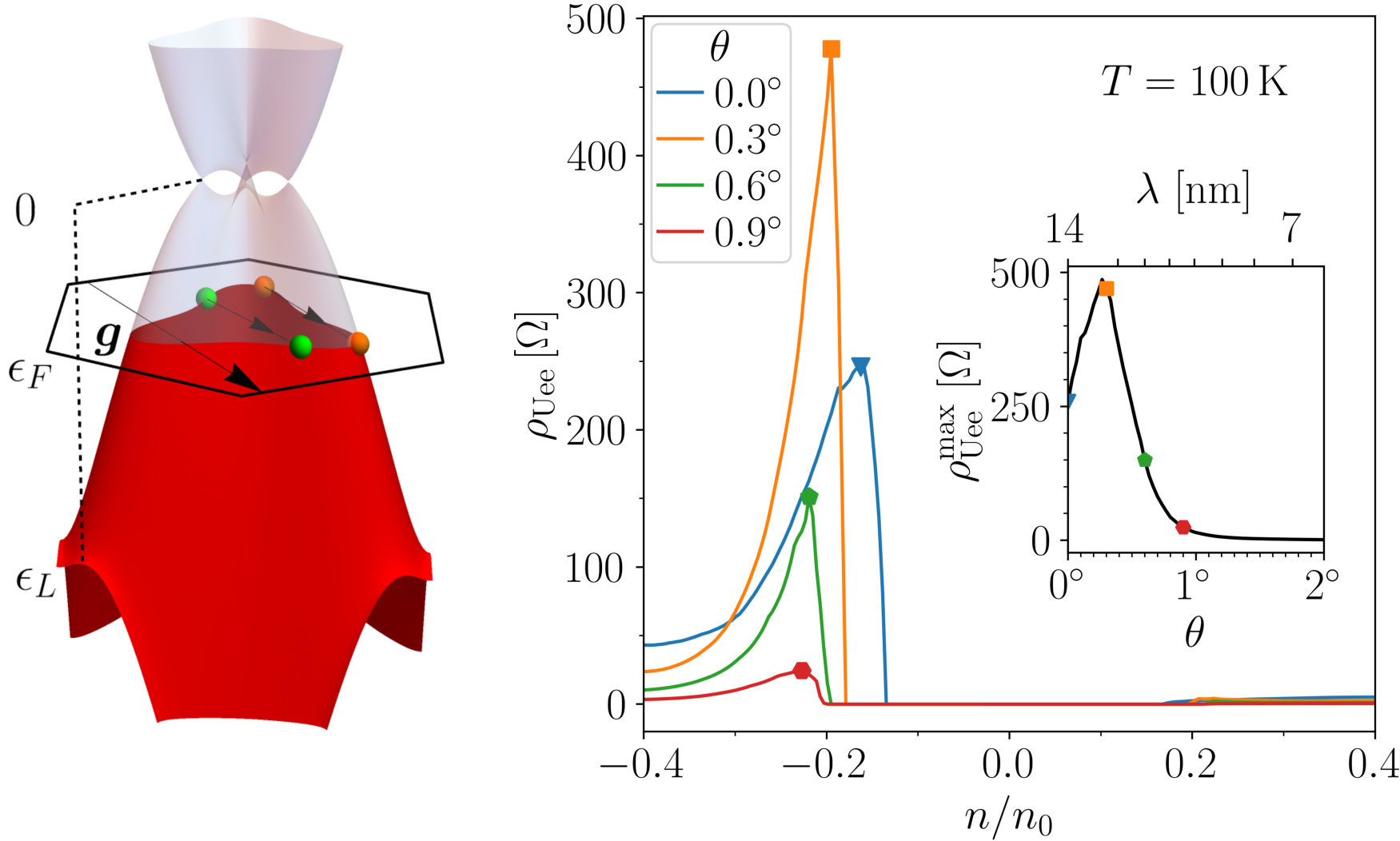}
\caption{Left: Umklapp electron-electron (Uee) scattering by a moir\'e superlattice in BLG. $\epsilon_F $ and $\epsilon_L $ are the Fermi energy and saddle point energy in the first mSL miniband on the valence side, respectively, counted from the conduction-valence band edge. Right: The non-monotonic evolution of the contribution, $\rho_\mathrm {Uee} = T^2 f(n)$, of Uee scattering to the electrical resistivity against electron density, $n$, for various twists angles, $\theta $, between graphene and hBN, at $ T =\SI {100} {K} $ (Uee processes dominate when $ T\ll | \epsilon_F |/k_B, | \epsilon_F - \epsilon_L |/k_B $). Inset: Peak value of the Uee resistivity, whose magnitude, $\rho_\mathrm {Uee} ^\mathrm {max} $, is shown as a function of the mSL period, $\lambda $ (and $\theta $).}
\label{fig: 1}
\end{figure}

Umklapp electron-electron (Uee) scattering is a fundamental process contributing towards the electrical resistivity of ultraclean metals. In this process, a pair of electrons interact via Coulomb repulsion and simultaneously transfer momentum, $\hbar\bm {g}$, to the crystalline lattice, where $\bm {g}$ is a reciprocal lattice vector (Bragg vector) of this lattice. Taking into account this momentum kick, the wavevectors of the incoming ($\bm {k}_{1/2} $) and outgoing ($\bm {k}_{3/4} $) electron states satisfy the following condition:
\begin{equation}
\label{eq: conservation}
\bm {k}_3+\bm {k}_4 =\bm {k}_1+\bm {k}_2+\bm {g}.
\end{equation}
When such a process relocates a pair of electrons across the Fermi surfaces, as illustrated in Fig.~\ref{fig: 1} (left-hand-side panel), the resulting two-electron back-scattering generates resistivity, in contrast to ``normal'' Coulomb scattering, which conserves the total momentum of the pair. The Uee contribution towards the resistivity typically has a $T^2$ temperature dependence~\cite{\ueeQuadraticTemperature}, but it is difficult to otherwise vary its strength in metals, where the electron density and a size of the Fermi surface are set by the material's chemistry, and the latter may not contain states that satisfy the condition in Eq.~(\ref{eq: conservation}).

With the availability of long-period superlattices, such as moir\'e superlattices (mSL) in incommensurate heterostructures of graphene~\cite{\mSL, \experimentMsl} or twisted graphene bilayers~\cite{\twistedBilayerGraphene}, it becomes feasible to vary the electron density across the range where Uee processes can be switched on/off and, then, its strength substantially varied. In a monolayer graphene/hexagonal boron nitride (hBN) heterostructure, it has been observed that, above a well-defined threshold density (which depends on the mSL period and, therefore, on the twist angle between graphene and hBN crystals), the rate of mSL-Uee gradually increases, becoming a dominant factor in the resistivity at room temperature~\cite{\ueeMonolayer}.

Here, we claim that mSL-Uee processes are important for understanding electronic transport in highly aligned bilayer graphene (BLG)/hBN heterostructures, where they make a substantial contribution, 
\begin{equation}
\label{eq: threshold scaling}
\rho_{\mathrm {Uee}} \propto T^2|n-n_*|^{1/2},
\end{equation}
towards the resistivity, Fig.~\ref{fig: 1}. As for the mSL in monolayer graphene, this contribution appears only above a threshold density, $ n_*$, growing rapidly just above the threshold. However, at higher densities and specifically for bilayer graphene, the resistivity falls off with the density of states, which results in a prominent peak, $\rho_{\mathrm {Uee}} ^\mathrm {max} $, in the density-dependent resistivity. The size of this peak increases non-monotonically with the mSL period, $\lambda $, (maximum value for $\lambda\approx\SI {13} {nm} $) as a result of the interplay between the trigonal warping of the dispersion of electrons in BLG~\cite{\trigonal, \minivalley} and the mSL periodicity. This contrasts with the mSL in monolayer graphene, where the Uee resistivity increases monotonically with both density and mSL period~\cite{\ueeMonolayer}, $ \rho_{\mathrm {Uee}} \propto T ^ 2| n - n_*| ^ {3/2}$, due to the suppressed backscattering of Dirac electrons.

\begin{figure}[ht!]
\centering
\vspace{10pt}
\includegraphics[width=\linewidth]{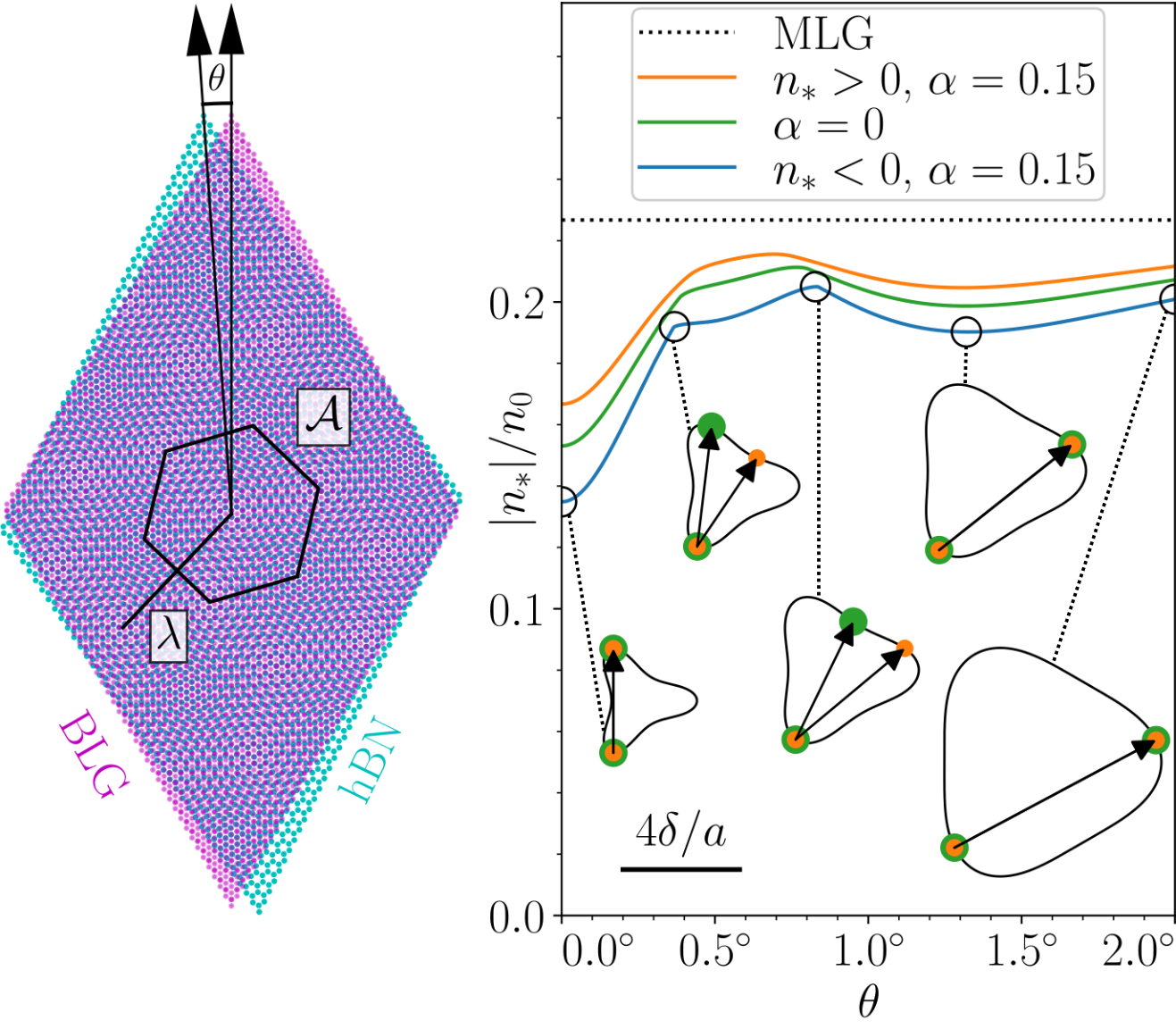}
\caption{Left: The lattice mismatch ($\delta \approx 1.8\% $) and twist, $\theta $, between BLG and hBN gives rise to a moir\'e superlattice with period, $\lambda \approx a/\sqrt{\delta ^ 2+\theta ^ 2}$, and unit cell of area, $\mathcal{A} =\sqrt{3}\lambda^2/2$. Right: The mSL-normalised magnitude, $| n_*|/n_0 $ ($ n_0 = 4/\mathcal {A} $) of the threshold density of holes, $n_*<0$, or electrons, $n_*>0$, at which Uee scattering becomes possible due to a sufficiently large Fermi line. The threshold density, $n_*$, was calculated as a function of twist angle $\theta$ taking into account the particle-hole asymmetry in the BLG Hamiltonian ($\alpha=0.15$), and compared to the symmetric cases of $\alpha=0$ and the monolayer graphene superlattice ($|n_*|\approx 0.23 n_0$).}
\label{fig: lattice}
\end{figure}

The above predictions are derived by considering Uee scattering in the BLG/hBN heterostructure sketched in Fig.~\ref{fig: lattice}, enabled by the mSL at the graphene/hBN interface, which period is determined by  a $\delta = 1.8$\% lattice mismatch between graphene and hBN and a misalignment angle, $\theta$. Projecting onto the low-energy bands of bilayer graphene in its $\bm {K} ^\xi $ valley ($\xi =\pm $), the electronic properties of this system can be described by a $ 2\times 2 $ effective Hamiltonian,~\cite{\bilayerHamiltonian, four-band,\zoneFolding}
\begin {widetext}
\begin{equation}
\label{eq: low-energy Hamiltonian}
\begin {gathered}
\hat {H} =\frac {- 1} {2 m_*}
\begin{pmatrix}
0 & \hat {\pi} ^{\dag ^ 2}\\
\hat {\pi} ^ 2 & 0
\end{pmatrix}
+ v_3
\begin{pmatrix}
0 & \hat {\pi} \\
\hat {\pi} ^\dag & 0
\end{pmatrix}
+\frac {\alpha\hat {\bm {p}} ^ 2} {2 m_*}
\begin{pmatrix}
1 & 0\\
0 & 1
\end{pmatrix}
-\frac {\Delta} {2}\bigg (1 -\frac {\hat {\bm {p}} ^ 2} {m_*\gamma_1}\bigg)
\begin{pmatrix}
1 & 0\\
0 & -1
\end{pmatrix}
+\sum_{m = 0} ^ 5\hat M_{\bm {g}_m}, 
\\
\hat M_{\bm {g}_m} = 
\begin{pmatrix}
[u_0+ i (-1) ^ m u_3] e ^ {i\bm {g}_m \cdot\bm {r}} &\frac 1 {\sqrt {2 m_*\gamma_1}}  u_1 (-1) ^ {m + 1} e ^ {- i\xi m \pi/3} e ^ {i\bm {g}_m \cdot\bm {r}} \hat\pi ^\dag\\
\frac 1 {\sqrt {2 m_*\gamma_1}}u_1 (-1) ^ {m + 1} e ^ {i\xi m \pi/3} \hat\pi e ^ {i\bm {g}_m \cdot\bm {r}} & \frac 1 {2 m_*\gamma_1} [u_0 - i (-1) ^ m u_3] \hat\pi e ^ {i\bm {g}_m \cdot\bm {r}}\hat\pi ^\dag
\end{pmatrix},
\end {gathered}
\end{equation}
\end {widetext}
where $\hat\pi = \hbar ( - i\xi\partial_x +\partial_y) $ and $\hat {\bm {p}} ^ 2 = -\hbar ^ 2 (\partial_x ^ 2+\partial_y ^ 2) $. The first three terms are intrinsic to BLG, representing the effective electron mass, $ m_* =\gamma_1/(2 v ^ 2) \approx 0.032\, m_e $, from the intralayer ($ v $) and vertical interlayer ($\gamma_1 $) couplings, trigonal warping from the skew interlayer ($ v_3 $) couplings~\cite{parameters} and a parabolic shift which lifts the particle-hole (eh) symmetry ($\alpha $)~\cite{alpha}, respectively. The fourth term represents an electrostatically controlled interlayer potential asymmetry, $\Delta $.

The final term in Eq.~(\ref{eq: low-energy Hamiltonian}) represents the effects of the mSL sketched in Fig.~\ref{fig: lattice}, with harmonics corresponding to the first star of mSL Bragg vectors, $\bm {g}_m\approx\delta \cdot\bm {G}_m -\theta (\bm {e}_z\times\bm {G}_m) $ ($ m = 0, 1, \cdots, 5 $), where $\bm {G}_m =\frac {4\pi} {\sqrt {3} a} (-\sin \frac {m\pi} 3, \cos \frac {m\pi} 3) $ is a graphene Bragg vector. This is parameterised by $ u_{0/1/3} $, corresponding to an energy shift, gauge field and mass term in the graphene layer closest to the hBN layer, respectively~\cite{superlattice_parameters}. Each harmonic, $\hat M_{\bm {g}_m } $, couples plane wave states separated by $\bm {g}_m $, which reconstructs the conduction and valence bands of isolated BLG into minibands (see Fig.~\ref{fig: 1}).

For electrons on a superlattice, Coulomb interaction leads to mSL-Uee processes (see Fig.~\ref{fig: 1}): two electrons from one side of the Fermi line backscatter together to the other side, receiving a momentum kick ($\bm {g} =\bm {g}_m $) from the mSL. Such processes only occur when the size of the Fermi contour is sufficiently large compared to $ |\bm {g}|$, giving a threshold electron density, $ n_*$, which decreases with the size of the mSL unit cell as seen in Fig.~\ref{fig: lattice} for the gapless spectrum with $\Delta = 0 $. The mSL-normalised threshold, $ n_*/ n_0 $, (due to spin-valley degeneracy, $n_0 = 4/\mathcal{A}$ corresponds to one filled miniband) increases non-monotonically with $\theta $ as the isoenergy lines become decreasingly concave with density, a consequence of the trigonal warping and most significant when $|n| <2 m ^ 2 v_3 ^ 2/\pi $. This distinguishes the threshold behaviour in bilayer graphene, pulling down from the value $| n_*|/ n_0 = \pi/(8\sqrt 3)\approx 0.23$ established for the the isotropic Dirac spectrum of monolayer graphene. Also, it is important to consider the effects of a conduction-valence band asymmetry in the bilayer dispersion, accounted for by the third term in Eq.~(\ref{eq: low-energy Hamiltonian}) with $\alpha = 0.15 $~\cite{parameters}. The latter affects the concavity of the isoenergy lines, especially in the first valence miniband, making the threshold density, $| n_*|$, slightly different for $n$- and $p$-doping of BLG mSL - see Fig.~\ref{fig: lattice}. 

In the following, we derive the amplitudes for mSL-Uee processes, treating the mSL and electron-electron interactions in the lowest-order perturbation scheme. This is implemented for densities just above the threshold, $ n_*$, where the resonant mixing of plane wave states is negligible, and we neglect the reconstruction of the electron dispersion into minibands. We account for the four leading Feynman diagrams involving Coulomb and mSL scattering of electrons off and back on to the Fermi level via an intermediate virtual state,
\begin {widetext}
\begin{equation}
\label{eq: amplitudes}
\includegraphics[width=\linewidth]{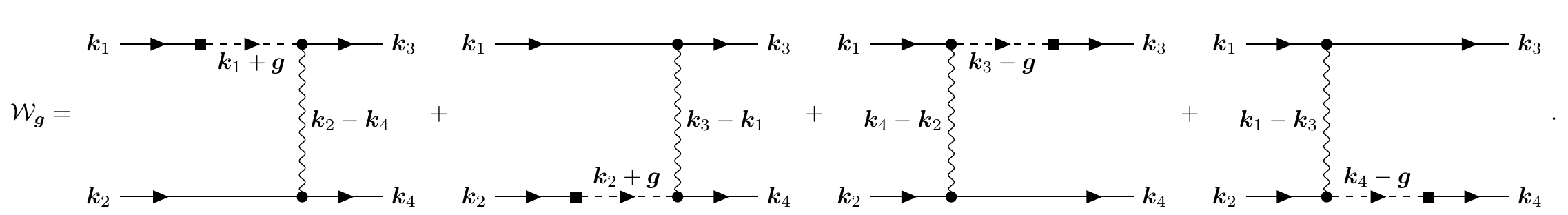}
\end{equation}
\end{widetext}
In each diagram, the initial and final momenta are related by Eq.~(\ref{eq: conservation}); \includegraphics{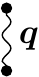}$\equiv \tilde V (\bm {q}) \approx W $ is the screened Coulomb interaction; $\blacksquare\equiv\hat M_{\bm {g}} $ is an mSL interaction harmonic which imparts momentum kick $\hbar\bm {g} $; and \includegraphics{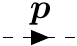} is a propagator of an electron in the virtual state~\cite{\screening, overlaps}. To mention, the mSL scattering amplitudes feature a particle-hole asymmetry, generic for graphene/hBN heterostructures, with values typically an order of magnitude larger in the valence miniband as compared to the conduction miniband~\cite{\mSL}.

Equipped with the amplitudes in Eq.~(\ref{eq: amplitudes}), we use linear transport theory~\cite{\electronElectronLinearTransport} (see SM) to calculate the contribution of Uee processes to the resistivity,
\begin {widetext}
\begin{equation}
\label{eq: Uee resistivity}
\rho_\mathrm {Uee} = \frac h {6 e ^ 2} (k_B T) ^ 2\sum_{m = 0} ^ 5 \int \frac {d\theta_1 d\theta_2} {|\bm {k}_3 \times \bm {k}_4|} \frac {k_1 k_2 k_3 k_4} {| v_{k1} v_{k2} v_{k3} v_{k4} |} |\mathcal {W}_{\bm {g}_m} | ^ 2 v_{x 1}( v_{x 1}+ v_{x 2}- v_{x 3}- v_{x 4}) \bigg/\bigg (\int d\theta\frac k {| v_k|} v_x ^ 2\bigg) ^ 2.
\end{equation} 
\end {widetext}
In this expression, $\bm {k}_i = k_i (\cos\theta_i,\sin\theta_i) $ is the wave vector of each electron ($ i = 1, 2, 3, 4 $) on the Fermi line, and $\bm {v}_i $ its group velocity. In Fig.~\ref{fig: 1}, the results of this analysis are summarised for the vertically unbiased heterostructure, $\Delta = 0 $. The Uee contribution is isotropic ($\rho_{\alpha\beta} ^\mathrm {Uee} \equiv \rho_\mathrm {Uee}\delta_{\alpha\beta} $) due to the $ C_3 $ symmetry of the mSL. Also, note that the ``normal'' (momentum-conserving) electron-electron scattering suppresses higher order harmonics in the non-equilibrium distribution of electrons, so that accounting for Uee becomes the same as accounting for an additional momentum transfer from the accelerated electrons (by the electric field) in the scattering time approximation (see SM). Here, we limit the analysis of Uee to the density range of $0.1n_0 < | n | <0.4n_0$, excluding from the analysis electron-hole scattering at the principal miniband edge and staying away from the mSL-induced van Hove singularity, where the perturbative treatment of the mSL interaction becomes inaccurate~\cite{\vanHoveResistivity}.

Typically, the Uee contribution in Eq.~(\ref{eq: Uee resistivity}), $\rho_\mathrm {Uee} \approx T^2 f(n)$, rises rapidly above the threshold, $f(n) \propto |n-n_*|^{1/2}$. This singular behaviour originates from the rapid expansion of the phase space around the incoming/outgoing points in Fig.~\ref{fig: lattice}, with $\bm {k}_3 \times \bm {k}_4 = 0 $ at the threshold. The exception is an initial interval of linear scaling, $f(n)\propto| n-n_*|$, found for the mSL with a twist angle in the range of $\SI {0.3} {\degree} <\theta <\SI {0.8} {\degree}$.

\begin{figure}
\centering
\includegraphics[width=\linewidth]{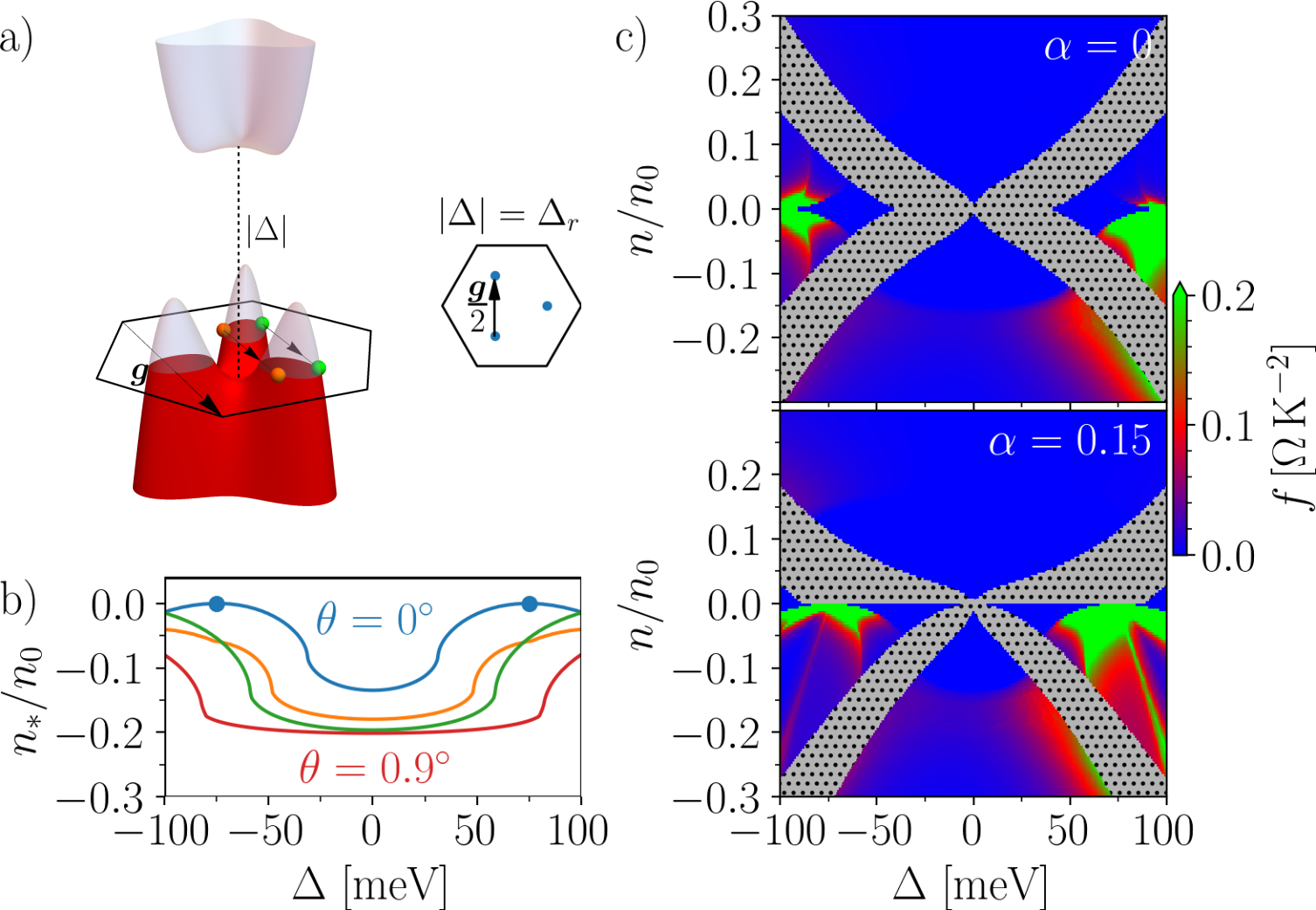}
\caption{a) Umklapp electron-election (Uee) scattering with momentum kick, $\hbar \bm {g} $, between the three minivalleys in the valence miniband for non-zero interlayer potential asymmetry, $\Delta$, opening a gap between the minibands. The minivalley edges are connected by $\bm {g}/2 $ when this system is aligned (zero twist, $\theta = 0 $) and $ |\Delta | =\Delta_r $ (inset). b) The non-monotonic evolution of the valence miniband threshold density, $ n_* $, with $\Delta $ for various twists, $ \theta = 0 ^\circ$ to $0.9 ^\circ$ from top to bottom, and $\alpha = 0.15$ ($ n_*= 0 $ when $\theta = 0 $ and $ |\Delta | =\Delta_r \approx\SI {75} {meV}$). c) The temperature-independent component, $ f $, of the dominant contribution, $\rho_\mathrm {Uee} \approx T ^ 2 f $, of Uee processes to the electrical resistivity against electron density, $ n $, and $\Delta $, with ($\alpha = 0 $) and without ($\alpha = 0.15 $) particle-hole symmetry, respectively. We exclude a (grey, dotted) butterfly-shaped region in each panel where the contribution of other processes are significant, whose wings are mirrored by zero layer polarisation ($\Delta = 0 $), and charge neutrality ($ n = 0 $) when $\alpha = 0 $.}
\label{fig: Delta}
\end{figure}

We also find that the interlayer potential asymmetry, $\Delta $, which opens a homogeneous bandgap in BLG~\cite{\tunableBandgap}, has a pronounced effect on the Uee processes. The gap promotes formation of three well-separated minivalleys at the BLG band edges, which persist up to the density $|n| \sim 2 m |\Delta | ^2/(\pi\gamma_1)$. The separation of the minivalleys increases with $|\Delta|$, thus, decreasing the threshold doping density, $|n_*|$, at which the Uee channel opens. For example, in the aligned BLG/hBN heterostructure ($\theta=0$) with $|\Delta|=\Delta_r \approx\SI {75} {meV}$, the minivalleys are separated by the $\frac12 \bm {g}_m $, so that Uee scattering transfers pairs of electrons between these minivalleys even at small doping (corresponding to $n_*= 0$, as shown in Fig. 3.b)).

The results of numerical computations of the Uee resistivity contribution, $\rho_\mathrm {Uee} $, across a broad range of parameters are summarised in Fig.~\ref{fig: Delta} c). We highlight the regions where the resistivity is dominated by Uee processes, excluding a butterfly-shaped region where thermally-activated electron-hole scattering processes may dominate. The wings of this butterfly, shown in the bottom panel, differ for $n$- and $p$-doping, which reflects the particle-hole asymmetry of the BLG dispersion (here, we use $\alpha = 0.15 $~\cite{alpha}), conversely being mirrored by charge neutrality ($n = 0$) for $\alpha = 0$ in the top panel. Regardless of $\alpha $, the wings are mirrored by $\Delta = 0$, where the wavefunctions feature zero layer polarisation. In contrast, the Uee contribution differs for positive and negative $\Delta$. This is because the interlayer asymmetry gap (vertical bias) shifts the weight of the low-energy electron states towards/away from the bottom graphene layer, hence, increasing/reducing the mSL scattering strength determined by the hBN crystal aligned with the BLG flake. Also, as in monolayer graphene mSL, Uee processes are much stronger for $p$-doping (first miniband of holes) than for $n$-doping, due to the particle-hole symmetry breaking by the mSL potential~\cite{PhysRevB.87.245408,Wallbank2019}

Overall, we predict a strong contribution of umklapp electron-electron scattering of moir\'e superlattice towards the resistivity of highly-aligned BLG/hBN heterostrucutres, with a non-monotonic density dependence near the Uee threshold. While the Uee role would increase at higher temperatures, at low densities (near the threshold) it will compete with electron-hole scattering processes, promoted by electron-hole activation across the conduction-valence band edge~\cite{\ehScattering}. Hence, for a more accurate description of the Coulomb scattering effect in the resistivity of mSL in bilayer graphene, one would need to account for both Uee and electron-hole scattering on equal footing. Also, one may want to extend the Uee analysis onto a broader range of miniband fillings, by calculating Uee rates using the full details of the mSL minibads spectra and Wannier functions, as attempted for a model graphene superlattice~\cite{Ishizuka_2022}. Such a calculation may offer additional features in $\rho_\mathrm {Uee} (n)$ when approaching the opposite (high-density) miniband edge or for deeper minibands, though, for quantititative validity, that has to be performed using a mSL model with the experimentally verified parameters~\cite{\mslPotentials}.

We thank A. Knothe, S. Slizovskiy, K. Novoselov and R. K. Kumar for useful discussions. We acknowledge support from EU Graphene Flagship Project, EPSRC Grants No. EPSRC CDT Graphene-NOWNANO EP/L01548X/1, EP/S019367/1, EP/P026850/1 and EP/N010345/1. All the research data supporting this publication is directly available within this publication and supplemental material accompanying this publication.

\bibliography{ref}% Produces the bibliography via BibTeX. 

\begin{thebibliography}{41}
\bibitem [{fou()}]{four-band}%
  \BibitemOpen
  \href@noop {} {}\bibinfo {note} {The full $ 4\times 4 $ Hamiltonian features
  two more bands at energies $\sim \pm\gamma_1 $ which are projected onto the
  lower-energy bands using a Schrieffer–Wolff
  transformation~\cite{\bilayerHamiltonian}.}\BibitemShut {Stop}%
\bibitem [{ove()}]{overlaps}%
  \BibitemOpen
  \href@noop {} {}\bibinfo {note} {We include the overlaps of the wavefunctions
  of the incoming and outgoing electrons with the operator at each
  vertex.}\BibitemShut {Stop}%
\end{thebibliography}

%%%%%%%%%% Merge with supplemental materials %%%%%%%%%%
\pagebreak
\onecolumngrid
\begin{center}
\textbf{\large Supplemental Material for ``\mytitle''}
\end{center}
%%%%%%%%%% Merge with supplemental materials %%%%%%%%%%
%%%%%%%%%% Prefix a "S" to all equations, figures, tables and reset the counter %%%%%%%%%%
\setcounter{equation}{0}
\setcounter{figure}{0}
\setcounter{table}{0}
\setcounter{page}{1}
\makeatletter
\renewcommand{\theequation}{S\arabic{equation}}
\renewcommand{\thefigure}{S\arabic{figure}}
\renewcommand{\bibnumfmt}[1]{[S#1]}
\renewcommand{\citenumfont}[1]{S#1}
%%%%%%%%%% Prefix a "S" to all equations, figures, tables and reset the counter %%%%%%%%%%

% \bibliography{ref}% Produces the bibliography via BibTeX. 

\section {Non-orthogonality of monolayer graphene Hamiltonian}
\label{sec: alpha}

The generic $ 2\times 2 $ Hamiltonian for the two sublattices of monolayer graphene, $ A $ and $ B $, in the $ (A, B) $ basis is
\begin{equation}
\hat H =
\begin{pmatrix}
\epsilon_A & -\gamma_0f (\bm {k})\\
-\gamma_0 f ^*(\bm {k}) &\epsilon_B
\end{pmatrix},
\end{equation}
where $\epsilon_{A/B} $ are the on-site energy of the $ 2 p ^ z $ graphene orbitals in each sublattice, $ A/B $, $\gamma_0 $ is the nearest-neighbour hopping parameter, and
\begin{equation}
f (\bm {k}) = e ^ {i k_y a/\sqrt 3} +2 e ^ {-i k_y a/(2\sqrt 3)}\cos\bigg (\frac {k_x a} 2\bigg),
\end{equation}
for plane wave states of momentum $\hbar\bm {k}~$\cite{nanotubes}. The $ 2 p ^ z $ orbitals on each sublattice are non-orthogonal, so that we derive the energy eigenvalues, $\epsilon $, and corresponding wavefunction, $\psi $, by solving the generalised eigenvalue equation,
\begin{equation}
\hat H\ket {\psi (\bm {k})} =\epsilon S \ket {\psi (\bm {k})},
\end{equation}
where the overlap matrix is
\begin{equation}
S =\begin{pmatrix}
1 & sf (\bm {k})\\
s f ^*(\bm {k}) & 1
\end{pmatrix},
\end{equation}
with $ s \approx 0.13 $ quantifying this overlap. We can orthogonalize this Hamiltonian using the transformation
\begin{equation}
\hat H\rightarrow S ^ {-1/2}\hat H S ^ {-1/2}\approx \hat H -\frac 1 2\{\hat H, S - I\},
\end{equation}
to first order in $ S $, where $ I $ is the $ 2\times 2 $ identity matrix, and $\{A, B\}\equiv AB + BA $ is an anti-commutator. This results in an extra term in the Hamiltonian, $ s\gamma_0 | f (\bm {k}) | ^ 2I $, which breaks particle-hole symmetry. 

\section {Effective two-band model of bilayer graphene}
\label{sec: two band} 

The full $ 4\times 4 $ Hamiltonian of bilayer graphene accounts for each of the two sublattices, $ A $ and $ B $, in each layer~\cite{\bilayerHamiltonian,\bilayerHamiltonianParameters}. However, in Bernal stacking, the $ A $ atoms of the top layer are directly above the $ B $ atoms of the bottom layer (see Fig.~2 of the main text) and have a large interlayer coupling, $\gamma_1 $. This results in two high-energy bands ($ |\epsilon | >\gamma_1 $) primarily located on these dimer atoms, alongside the low-energy bands primarily located on the non-dimer atoms which dominate the contribution of Umklapp electron-electron (Uee) processes to the resistivity.

We use a Schrieffer-Wolff transformation to project the high-energy bands onto the low-energy bands, giving an effective $ 2\times 2 $ Hamiltonian at the $\bm {K} ^ \pm $ points:
\begin{equation}
\label{eq: full low-energy Hamiltonian}
\hat {H}_0 =\frac {- 1} {2 m_*}
\begin{pmatrix}
0 & \hat {\pi} ^{\dag ^ 2}\\
\hat {\pi} ^ 2 & 0
\end{pmatrix}
+ v_3
\begin{pmatrix}
0 & \hat {\pi} \\
\hat {\pi} ^\dag & 0
\end{pmatrix}
+\frac {\alpha_1} {2 m_*}
\begin{pmatrix}
\hat {\pi} ^\dag \hat {\pi} & 0\\
0 & \hat {\pi}\hat {\pi} ^\dag
\end{pmatrix}
-\frac {\Delta} {2}\bigg [
\begin{pmatrix}
1 & 0\\
0 & -1
\end{pmatrix}
-\frac 1 {m_* \gamma_1}
\begin{pmatrix}
\hat {\pi} ^\dag \hat {\pi} & 0\\
0 & -\hat {\pi}\hat {\pi} ^\dag
\end{pmatrix}
\bigg].
\end{equation}
Note that we neglect a constant energy term corresponding to the mean on-site potential. Particle-hole symmetry breaking is achieved by the third term ($\alpha_1\approx 0.15$), alongside a small, additional parabolic shift, $ \frac {\alpha_2\hat {\bm {p}} ^ 2} {2 m_*} I$ ($\alpha_2\sim 0.01 $), which accounts for the the next-nearest-neighbour intralayer couplings and non-orthogonality of graphene orbitals in Sec.~\ref {sec: alpha}, expanded at the $\bm {K} ^ \pm $ points. The momentum operators commute in the absence of an external magnetic field, and Eq.~(\ref{eq: full low-energy Hamiltonian}) simplifies to give the first four terms in Eq.~(3) of the main text, with $\alpha =\alpha_1+\alpha_2 \approx 0.15 $~\cite{parameters,alpha}. Considering plane wave states of wave vector, $(\bm {K} ^ \pm +)\bm {k} $, in the Brillouin zone, this Hamiltonian features a conduction and valence band of energies, $\epsilon_\pm (\bm {k})$ ($\epsilon_+ >\epsilon_- $), and wavefunctions, $\ket {\psi_\pm (\bm {k})} = e ^ {i\bm {k} \cdot\bm {r}} u_\pm (\bm {k}) $, respectively, satisfying the eigenvalue equation, $\hat H_0 \ket {\psi_\pm (\bm {k})} = \epsilon_\pm (\bm {k}) \ket {\psi_\pm (\bm {k})}$.

The final term in the Hamiltonian in Eq.~(3) of the main text represents the moir\'e superlattice (mSL) interaction with the hBN layer, featuring harmonics, $\hat M_{\bm {g}_m} $, for each of the first star of mSL Bragg vectors, $\bm {g}_m $ ($ m = 0, 1, \cdots 5 $)~\cite{\mSL,\mslPotentials}. The harmonic $\hat M_{\bm {g}_m} $ couples plane wave states of wavevector $\bm {k} $ and $\bm {k} +\bm {g}_m $: $\braket {\psi_{s'} (\bm {k} +\bm {g}_m) | \hat M_{\bm {g}_m} | \psi_s (\bm {k})} \neq 0 $ ($ s, s' =\pm $). Hence, we derive the reconstructed dispersion from the eigenvalue equation with the zone-folded wavefunction,
\begin{equation}
\label{eq: zone-folded wavefunction}
\ket {\Psi (\bm {k})} =\sum_{\bm {g}}\sum_{s =\pm} c_{\bm {k} +\bm {g}} ^ s \ket {\psi_s (\bm {k} +\bm {g})},
\end{equation}
where $ c_{\bm {k} +\bm {g}} ^ s $ are complex coefficients and we sum over the mSL Bragg vectors, $\bm {g} $~\cite{\zoneFolding}. This gives a collection of minibands (see Fig.~1) in the moir\'e Brillouin zone (mBZ), and the conduction and valence minibands are converged when we sum over mSL Bragg vectors, $\bm {g} $, which are the sum of at most two of the first star vectors, $\bm {g}_m $. As discussed in the main text, we can neglect the mSL-induced reconstruction of the bands in the relevant density ranges, $ n/n_*\ll 4 $.

Depending upon the interlayer potentially asymmetry, $\Delta $, and density, $ n $, the Fermi line will have one of three forms:
\begin{enumerate}
    \item A single contour centred on the origin.
    \item Two non-touching contours, each centred on the origin. We neglect this regime, since it is narrow on the energy axis, and the resistivity will be dominated by impurity scattering at the van Hove singularities~\cite{\vanHoveResistivity}.
    \item Three non-touching contours (minivalleys), $ p = 0, 1, 2 $, centred on the three band edges, $\bm {k}_c ^ {(p)} = k_c (\cos\frac {p 2\pi} 3, \sin\frac {p 2\pi} 3) $ ($ k_c >0 $), respectively~\cite{\minivalley}. The contours do not enclose the origin, and are related by the three-fold rotational symmetry. We expand about the centre of the minivalley, $\bm {k} \rightarrow \bm {k}_c^ {(p)} +\bm {k} $:
\begin{equation}
\bm {k} = k \bigg (\cos\bigg (\theta +\frac {p 2\pi} 3\bigg), \sin\bigg (\theta +\frac {p 2\pi} 3\bigg)\bigg),
\end{equation}
such that $ k (\theta) $ is the same for each minivalley. We suppress the contour index for simplicity in the following sections, implicitly summing over the valid contours.
\end{enumerate}

\section {Screened Coulomb interaction}
\label{sec: screening}

The unscreened Coulomb potential in real space is
\begin{equation}
V_0 (\bm {r}) \equiv V_0 (r) =\frac {e ^ 2} {4\pi\kappa\epsilon_0 r},
\end{equation}
where $\kappa\approx 2.5 $ is the dielectric constant of bilayer graphene in hBN~\cite{\hbnDielectricConstant}. Performing the in-plane Fourier transformation on this potential gives
\begin{equation}
\label{eq: unscreened potential}
\tilde V_0 (\bm {Q}) \equiv \tilde V_0 (Q) = \frac {e ^ 2} {2\kappa\epsilon_0 Q},
\end{equation}
for the momentum transfer, $\hbar\bm {Q} $. Electron states screen this potential and, in the random phase approximation~\cite{\screening}, the screened potential is given by
\begin{equation}
\tilde V (\bm {Q}) =\frac {\tilde V_0 (Q)}{1+\tilde V_0 (Q) \Pi_0 (\bm {Q})},
\end{equation}
where the static polarisation, including the two-fold spin degeneracy, is given by
\begin{equation}
\label{eq: polarisation}
\Pi_0 (\bm {Q}) = - 2\sum_{s, s'=\pm }\sum_{\xi =\pm} \int \frac {d ^ 2 k} {(2\pi) ^ 2} \frac {f_0(\epsilon_s (\bm {k}))- f_0 (\epsilon_{s'} (\bm {k} +\bm {Q}))}{\epsilon_s (\bm {k})- \epsilon_{s'} (\bm {k} +\bm {Q})} |\braket {\psi_s (\bm {k}) |\psi_{s'} (\bm {k} +\bm {Q})} | ^ 2,
\end{equation}
in terms of the equilibrium distribution function for Fermi energy, $\epsilon_F $,
\begin{equation}
\label{eq: equilibrium distribution}
f_0 (\epsilon) =\frac {1} {e ^ {(\epsilon -\epsilon_F) / (k_B T)} +1}.
\end{equation}
In the low-temperature regime discussed in the main text, we approximate the distribution function as a step, $\partial f_0/\partial\epsilon\approx -\delta (\epsilon -\epsilon_F) $, and the polarisation is dominated by intra-band overlaps and small momentum transfers, $\bm {Q} \approx 0 $:
\begin{equation}
\label{eq: zero polarisation}
\Pi_0 (\bm {Q})\approx \Pi_0 (0) = \frac 1 {\pi ^ 2} \int d\theta \frac k {| v_k |},
\end{equation}
where we integrate over the Fermi surface, and the electron group velocity is $\bm {v} (\bm {k}) =\hbar ^ {-1}\nabla_{\bm {k}}\epsilon (\bm {k}) $. Since the screening is strong ($\tilde V_0 (Q) \Pi_0 (\bm {Q})\gg 1 $) in the region of interest, we approximate the screened potential as a contact potential, $ V (\bm {r}) \approx W\delta (\bm {r}) $, where
\begin{equation}
\label{eq: contact potential}
W = \Pi_0 (0) ^ {-1}.
\end{equation}

\section {Matrix elements}
\label{sec: matrix elements}

The matrix element, $ \mathcal {W}_{\bm {g}} $, for Uee scattering by the mSL Bragg vector, $\bm {g} =\bm {g}_m $, is the sum of four diagrams where electrons 1, 2, 3 and 4 scatter off the mSL, respectively, shown as Feynman diagrams in Eq.~(4) of the main text. These give the respective terms in the explicit expression,
\begin{equation}
\label{eq: explicit matrix elements}
\mathcal {W}_{\bm {g}} 
= X_{\bm {g}} (1, 2, 3, 4)
+ X_{\bm {g}} (2, 1, 4, 3)
+X_{-\bm {g}} ^* (3, 4, 1, 2)
+X_{-\bm {g}} ^* (4, 3, 2, 1),
\end{equation}
where
\begin{equation}
\label{eq: matrix element X}
X_{\bm {g}} (a, b, c, d) = \sum_{s' =\pm} \frac {\braket{\psi_{s'} (\bm {k}_a+\bm {g}) | \hat {M}_{\bm {g}} |\psi_s (\bm {k}_a)} } {\epsilon_s (\bm {k}_a) -\epsilon _{s'} (\bm {k}_a+\bm {g})} \tilde V (\bm {k}_b - \bm {k}_d) \braket{\psi_s (\bm {k}_c) | \psi_{s'} (\bm {k}_a+\bm {g})} \braket{\psi_s (\bm {k}_d) | \psi_s (\bm {k}_b)},
\end{equation}
summing over both bands for the intermediate virtual electron~\cite{\ueeMonolayer}. Spatial inversion is equivalent to taking the complex conjugate of the matrix elements, which leaves the magnitude of Eq.~(\ref{eq: matrix element X}) unchanged. In the main text, we restrict the integral to the Fermi surface, $\epsilon_{1/2/3/4} =\epsilon_F $.

\section {Linear response theory}
\label{sec: linear response theory}

The Boltzmann equation for the electron distribution function, $ f (\bm {k}) $, of an electron with wave vector, $\bm {k} $, in the presence of an external electric field, $\bm {E} $, is given by
\begin{equation}
\label{eq: Boltzmann}
e \bm {E} \cdot\nabla_{\bm {k}} f (\bm {k}) = I \{ f (\bm {k}) \},
\end{equation}
where the collision integral, $ I \{ f (\bm {k})\}$, is determined by electron scattering~\cite{\electronElectronLinearTransport}. Note that the electron charge is negative, $ e <0 $. The equilibrium distribution (\ref {eq: equilibrium distribution}), appropriate for $\bm {E} = 0 $, satisfies the detailed balance condition, $ I\{ f_0 (\epsilon) \} = 0 $. We expand about this equilibrium distribution to first order in the chemical potential shift, $\phi (\bm {k}) $:
\begin{equation}
\label{eq: distribution}
f (\bm {k}) = f_0 (\epsilon (\bm {k})) -\frac {\partial f_0} {\partial\epsilon} \bigg|_{\epsilon = \epsilon (\bm {k})} \phi (\bm {k}).
\end{equation}
Expanding Eq.~(\ref{eq: Boltzmann}) to first order in $\phi (\bm {k}) $ gives the linearised Boltzmann equation,
\begin{equation}
\label{eq: linearised Boltzmann}
e E v_E (\bm {k}) \frac {\partial f_0} {\partial\epsilon} \bigg|_{\epsilon = \epsilon (\bm {k})} = I\{ \phi (\bm {k}) \},
\end{equation}
which relates the kinetic function, $ v_E (\bm {k}) =\bm {v} (\bm {k})\cdot \bm {E}/E $, to $\phi (\bm {k}) $. This equation is inverted to give the corresponding shift, $\phi (\bm {k}) $, for the known kinetic function, $ v_E (\bm {k})$, from which we calculate the scattering-limited longitudinal electrical conductivity as 
\begin{equation}
\label{eq: conductivity}
\sigma = \frac {4e} E \int \frac {d ^ 2\bm {k}} {(2\pi) ^ 2} v_E \frac {\partial f_0} {\partial\epsilon} \phi,
\end{equation}
with corresponding resistivity, $\rho\equiv\sigma ^ {-1} $. As a result of the $ C_3 $ symmetry, the resistivity is isotropic and independent of the field direction, so we set $\bm {E} = (E, 0, 0) $ for simplicity. 

The collision integral for Uee scattering is given by~\cite{\electronElectronLinearTransport,Kim2020}
\begin{equation}
\label{eq: umklapp collision integral}
I_\mathrm {Uee} \{ \phi (\bm {k}_1) \} = \frac {4} {k_B T} \sum_{m = 0} ^ 5\int \frac {d ^ 2\bm {k}_2d ^ 2\bm {k}_3d ^ 2\bm {k}_4} {(2\pi) ^ 6} 2\pi\delta (\Delta\epsilon) (2\pi) ^ 2\delta ^ {(2)} (\Delta\bm {k} -\bm {g}_m) |\mathcal {W}_{\bm {g}_m} | ^ 2 \Delta\phi \frac {1} {16}\prod_{i = 1}^{4} \mathrm {sech} \bigg( \frac {\epsilon_i -\epsilon_F} {2 k_B T} \bigg),
\end{equation}
in terms of the change,
\begin{equation}
\Delta f =f (\bm {k}_3) +f (\bm {k}_4) -f (\bm {k}_1) -f (\bm {k}_2),
\end{equation}
of the total $ f =\phi,\epsilon,\bm {k}$ during the process. The Dirac delta functions ensure energy and momentum (including kick $\hbar\bm {g} $) conservation of the incoming and outgoing electrons. At low temperatures, we restrict this to an integral over the Fermi surface since 
\begin{equation}
\delta (\epsilon_3+\epsilon_4 -\epsilon_1 -\epsilon_2 )\prod_{i = 1} ^ 4 \mathrm {sech} \bigg( \frac {\epsilon_i -\epsilon_F} {2 k_B T} \bigg) \approx \frac {32\pi ^ 2 (k_B T) ^ 3} 3\prod_{i = 1} ^ 4\delta (\epsilon_i -\epsilon_F ).
\end{equation}

We simplify using the constant relaxation time approximation, where the collision integral, $ I\{f (\bm {k})\} $, of a scattering process is approximated by
\begin{equation}
I\{f (\bm {k})\} = -\hbar\frac {f (\bm {k}) - f_0 (\epsilon (\bm {k}))} {\tau},
\end{equation}
in terms of a constant scattering time, $\tau $~\cite{\ueeMonolayer}. Expanding to first order in the chemical potential shift, $\phi (\bm {k}) $, the collision integral simplifies to
\begin{equation}
I\{\phi (\bm {k})\} = \frac \hbar\tau \frac {\partial f_0} {\partial\epsilon}\bigg |_{\epsilon=\epsilon (\bm {k})} \phi (\bm {k}).
\end{equation}
Inserting this into the linearised Boltzmann equation in Eq.~(\ref{eq: linearised Boltzmann}), we easily find the solution,
\begin{equation}
\phi (\bm {k}) = - e E \tau v_x (\bm {k}) /\hbar,
\end{equation}
with scattering time,
\begin{equation}
\frac \hbar\tau =\int d ^ 2 k\, I\{v_x\} v_x \bigg/\bigg (\int d ^ 2 k \frac {\partial f_0} {\partial\epsilon} v_x ^ 2 \bigg),
\end{equation}
found self-consistently for a collision integral linear in $\phi $. Then, the Uee contribution to the resistivity, $\rho_\mathrm {Uee}$, in the low-temperature regime is given by Eq.~(5) of the main text.

\section {Low-density resistivity}
\label{sec: low-density}

At density, $ n $, just above the threshold, $ n_*$, Uee scattering will be restricted to the immediate vicinity of the points shown in Fig.~2 of the main text. The wavevector, $\bm {k} $, velocity, $\bm {v} $, and scattering amplitudes, $\mathcal {W}_{\bm {g}} $, are approximately constant in this range, so the resistivity has the approximate form,
\begin{equation}
\rho_\mathrm {Uee}\propto T ^ 2 \sum_{m = 0} ^ 5\int \frac {d\theta_1 d\theta_2} {|\sin (\theta_3 -\theta_4)|}.
\end{equation}
The size of the phase space in the integral scales according to $ | n - n_*| $, so we have $\rho_\mathrm {Uee}\propto T ^ 2| n - n_*| ^ {1/2} $, except for $\rho_\mathrm {Uee}\propto T ^ 2| n - n_*| $ in the region discussed in the main text corresponding to split Uee scattering, where $ \sin (\theta_3 -\theta_4)$ is also approximately constant.

\end{document}